\documentclass[twocolumn]{aastex63}

\usepackage{epstopdf}
\usepackage{mathtools}
\usepackage{mathrsfs}
\usepackage{calrsfs}
\usepackage{multirow}
\usepackage{xspace}
\usepackage[hang,flushmargin]{footmisc}
\usepackage[autostyle, english = american]{csquotes}
\usepackage{eucal}
\usepackage{amsmath}
\usepackage{url}

\usepackage{bm}

\newcommand{\V}{{\bm V}}
\newcommand{\D}{{\bm D}}
\newcommand{\G}{{\bm G}}
\newcommand{\I}{{\bm I}}
\newcommand{\R}{{\bm R}}
\newcommand{\Par}{{\bm P}}
\newcommand{\J}{{\bm J}}

\newcommand{\jp}[1]{\textbf{#1}}

\received{--}
\revised{--}
\accepted{--}

\submitjournal{ApJ}

\shortauthors{Park et al.}

\graphicspath{{./}}

\begin{document}

\title{Calibrating VLBI Polarization Data Using GPCAL. I. Frequency-Dependent Calibration}

\correspondingauthor{Jongho Park}
\email{jpark@kasi.re.kr}

\author[0000-0001-6558-9053]{Jongho Park}
\affiliation{Department of Astronomy and Space Science, Kyung Hee University, 1732, Deogyeong-daero, Giheung-gu, Yongin-si, Gyeonggi-do 17104, Republic of Korea}

\affiliation{Korea Astronomy and Space Science Institute, Daedeok-daero 776, Yuseong-gu, Daejeon 34055, Republic of Korea}
\affiliation{Institute of Astronomy and Astrophysics, Academia Sinica, P.O. Box 23-141, Taipei 10617, Taiwan}

\author[0000-0001-6988-8763]{Keiichi Asada}
\affiliation{Institute of Astronomy and Astrophysics, Academia Sinica, P.O. Box 23-141, Taipei 10617, Taiwan}

\author[0000-0003-1157-4109]{Do-Young Byun}
\affiliation{Korea Astronomy and Space Science Institute, Daedeok-daero 776, Yuseong-gu, Daejeon 34055, Republic of Korea}
\affiliation{University of Science and Technology, Gajeong-ro 217, Yuseong-gu, Daejeon 34113, Republic of Korea}

\begin{abstract}

In this series of papers, we present new methods of frequency- and time-dependent instrumental polarization calibration for Very Long Baseline Interferometry (VLBI). In most existing calibration tools and pipelines, it has been assumed that instrumental polarization is constant over frequency within the instrument bandwidth and over time. The assumption is not always true and may prevent an accurate calibration, which can result in degradation of the quality of linear polarization images. In this paper, we present a method of frequency-dependent instrumental polarization calibration that is implemented in GPCAL, a recently developed polarization calibration pipeline. The method is tested using simulated data sets generated from real Very Long Baseline Array (VLBA) data. We present the results of appyling the method to real VLBA data sets observed at 15 and 43 GHz. We were able to eliminate significant variability in cross-hand visibilities over frequency that is caused by frequency-dependent instrumental polarization. As a result of the calibration, linear polarization images were slightly to modestly improved as compared to those obtained without frequency-dependent instrumental polarization calibration. We discuss the reason for the minor impact of frequency-dependent instrumental polarization calibration on existing VLBA data sets and prospects for applying the method to future VLBI data sets, which are expected to provide very large bandwidths.

\end{abstract}

\keywords{high angular resolution --- techniques: interferometric --- techniques: polarimetric --- methods: data analysis}

\section{Introduction} \label{sec:intro}

The Very Long Baseline Interferometry (VLBI) technique is a method for obtaining very high angular resolution by using telescopes that are widely separated. The telescopes in a VLBI array can split the sky signal into two orthogonal polarization channels, which are typically right and left circularly polarized (RCP and LCP, respectively), and each can record independently. The observed polarized visibilities, which are the time average of the complex correlations between the electric fields for each polarization incident at each telescope in the array, are directly related to the distributions of the four Stokes parameters in the sky by means of the van Cittert-Zernike theorem \citep{vanCittert1934, Zernike1938, Thompson2017}. 

The relationship between the measured polarimetric visibilities and the source's intrinsic polarization can be distorted due to imperfections in the instrument's response. This distortion is primarily caused by antenna polarization that is not perfectly circular, which results in ``instrumental polarization" in the visibility data. Instrumental polarization is commonly described in terms of an antenna's response to a wave of polarization orthogonal to the nominal antenna response \citep[e.g.,][their Chapter 4.7]{Thompson2017}, which is often referred to as ``polarization leakages" or ``D-terms".

LPCAL, a task implemented in the Astronomical Image Processing System \citep[AIPS,][]{Greisen2003} based on the linearized leakage model \citep{Leppanen1995}, has been a standard method for calibrating VLBI data for nearly three decades. This method has been successful in a variety of studies using a variety of VLBI arrays \citep[e.g.,][]{Gomez2016, Casadio2017, Jorstad2017, Lister2018, Park2018, Park2019b, EHT2021a}. However, LPCAL has some limitations that prevent accurate calibration. It assumes that the total intensity and linear polarization structures of a calibrator are similar, and also it cannot fit the instrumental polarization model to data from multiple calibrators at the same time. The former is referred to as the ``similarity approximation"\footnote{As far as we are aware, the term "similarity approximation" was first introduced in \cite{Leppanen1995}. The original idea was proposed by \cite{Cotton1993}, who also cautioned that the assumption of a direct proportionality between total intensity and linear polarization emission may not hold.} \citep{Cotton1993, Leppanen1995}, and it is generally violated for VLBI data, especially at high frequencies at which nearly all calibrators are resolved.

The Generalized Polarization CALibration pipeline (GPCAL; \citealt{Park2021a}) has been developed to overcome these limitations\footnote{Other calibration/imaging pipelines that have recently been developed are \texttt{polsolve} \citep{MartiVidal2021}, similar to GPCAL but based on CASA \citep{Janssen2019, CASA2022, vanBemmel2022}, the \texttt{eht-imaging} software library using the regularized maximum likelihood technique \citep{Chael2016, Chael2018, Chael2022}, D-term Modeling Code (DMC, \citealt{Pesce2021}), THEMIS \citep{Broderick2020}, and Comrade \citep{Tiede2022} using Markov chain Monte Carlo schemes. See \cite{PA2022} for more details.} \citep{Park2021a}. The GPCAL pipeline is written in ParselTongue \citep{Kettenis2006} and is based on AIPS and the Caltech Difmap package \citep{Shepherd1997}. It enables the use of more accurate linear polarization models of calibrators for D-term estimation without being limited by the similarity approximation. It can also fit the instrumental polarization model to data from multiple calibrator sources simultaneously in order to increase the accuracy of the fitting.

GPCAL was applied to the polarization analysis of the first M87 Event Horizon Telescope (EHT) results \citep{EHT2021a} as well as other recent studies \citep[e.g.,][]{Takamura2023}. We have also demonstrated that GPCAL is capable of detecting polarizations from sources that are very weakly polarized \citep{Park2021c}. The linear polarization structure of the subparsec core of the M87 jet observed with the Very Long Baseline Array (VLBA) at 43 GHz is compact and has a very low fractional polarization ($\sim0.2$-$0.6\%$). It was not possible to well constrain this structure in previous studies with LPCAL using the same data set due to a breakdown in the similarity approximation, resulting in less accurate calibration of the instrumental polarization \citep{Walker2018, Kravchenko2020}.

In GPCAL, one assumption was made that can limit the accuracy of instrumental polarization calibration: it assumes that polarimetric leakages remain constant across the frequency bandwidth of a receiver and over time. The assumption of constant leakages over frequency may not apply to many VLBI arrays that have offered a wide bandwidth in recent years. In general, the magnitudes of leakages tend to increase as the operating frequency deviates further from the nominal frequency within the band. This is due to the design of the instruments, including polarizers, which are optimized to minimize leakages at the nominal frequency \citep[e.g.,][]{HP2015}. Based on recent wide-bandwidth VLBA data, some previous studies have suggested that frequency-dependent D-terms can limit the dynamic range of linear polarization images \citep{Kravchenko2020}. Also, the assumption of constant leakages over time is incorrect since polarimetric leakages are not constant across telescope beams; they are direction-dependent \citep{Smirnov2011}. The effect is more severe for large telescopes and at high frequencies, where antenna pointing is usually inaccurate.

In this series of papers, we present GPCAL implementations of novel methods of frequency and time-dependent instrumental polarization calibration of VLBI data. In this paper (Paper I hereafter), we present a method for calibrating frequency-dependent instrumental polarization. An accompanying publication (Park et al. 2023; Paper II) describes a method of calibrating instrumental polarization that varies with time. These methods are applied to real VLBI data sets and demonstrate that, depending on the circumstances, they can significantly improve calibration accuracy.

This paper is organized as follows. In Section~\ref{sec:model}, we explain the radio interferometer measurement equation, which GPCAL is based on to model the instrumental polarization. In Section~\ref{sec:pipeline}, we introduce the calibration procedure for frequency-dependent leakage calibration implemented in GPCAL. In Section~\ref{sec:synthetic}, we validate the method using synthetic data. In Section~\ref{sec:application}, we present results of applying the method to real observations from the VLBA at 15 and 43 GHz. We conclude in Section~\ref{sec:conclusion}.

\section{The Radio Interferometer Measurement Equation} 
\label{sec:model}

The observed polarized visibilities on a baseline between stations $m$ and $n$ are given by the correlation products
\begin{eqnarray}
\label{eq:product}
r^{RR}_{mn} &\equiv& \langle E_{R,m}E^*_{R,n}\rangle \nonumber\\
r^{LL}_{mn} &\equiv& \langle E_{L,m}E^*_{L,n}\rangle \nonumber\\
r^{RL}_{mn} &\equiv& \langle E_{R,m}E^*_{L,n}\rangle \nonumber\\
r^{LR}_{mn} &\equiv& \langle E_{L,m}E^*_{R,n}\rangle,
\end{eqnarray}
where $E$ is an electric field, $R$ indicates RCP, $L$ indicates LCP, angular brackets denote a time average, and an asterisk denotes complex conjugate. One can represent the visibilities using a coherency matrix formalism:

\begin{equation}
\V_{mn} = 
\begin{pmatrix}
r^{RR}_{mn} & r^{RL}_{mn}\\
r^{LR}_{mn} & r^{LL}_{mn}
\end{pmatrix}.
\end{equation}
The observed $\V_{mn}$ are corrupted by antenna gains and polarimetric leakages. The relationship between the true $\bar\V_{mn}$ and the observed $\V_{mn}$ can be described via a sequence of linear transformations, which is the so-called radio interferometer measurement equation (RIME; \citealt{Hamaker1996}, see also \citealt{Sault1996, HB1996, Hamaker2000, Smirnov2011}),
\begin{equation}
\label{eq:rime}
    \V_{mn} = \J_m \bar\V_{mn} \J^H_n,
\end{equation}
where H is the Hermitian operator and $\J$ is the so-called Jones matrix \citep{Jones1941}:
\begin{eqnarray}
    \J_{m} &=& \G_m \D_m \Par_m \nonumber\\
    &=& 
    \begin{pmatrix}
    G^R_m & 0 \\
    0 & G^L_m
    \end{pmatrix}
    \begin{pmatrix}
    1 & D^R_m \\
    D^L_m & 1
    \end{pmatrix}
    \begin{pmatrix}
    e^{-j\phi_m} & 0 \\
    0 & e^{+j\phi_m}
    \end{pmatrix}.
\end{eqnarray}
$G$ is the complex antenna gain, $D$ is the leakage factor (``D-term"), and $\phi$ is the antenna field rotation angle. Subscripts denote antenna numbers, and superscripts denote polarization. The field rotation angle takes the form
\begin{equation}
    \phi = f_{\rm el}\theta_{\rm el} + f_{\rm par}\psi_{\rm par} + \phi_{\rm off},
\end{equation}
where $\theta_{\rm el}$ is the source's elevation, $\psi_{\rm par}$ the parallactic angle, and $\phi_{\rm off}$ is a constant offset for the rotation of antenna feed with respect to the azimuth axis. Cassegrain mounts have $f_{\rm par} = 1$ and $f_{\rm el} = 0$ and thus the field rotation angle is equivalent to the parallactic angle, except for the constant offset. Nasmyth mounts have $f_{\rm par} = 1$ and $f_{\rm el} = +1$ for Nasmyth-Right type and $f_{\rm par} = 1$ and $f_{\rm el} = -1$ for Nasmyth-Left type.

For circular feeds, the $\bar\V$ are related to the Fourier transforms of the Stokes parameters ($\tilde{I}$, $\tilde{Q}$, $\tilde{U}$, and $\tilde{V}$) via
\begin{equation}
\label{eq:stokes}
\bar\V_{mn} \equiv
\begin{pmatrix}
\mathscr{RR} & \mathscr{RL} \\
\mathscr{LR} & \mathscr{LL}
\end{pmatrix}
=
\begin{pmatrix}
\tilde{I}_{mn} + \tilde{V}_{mn} & \tilde{Q}_{mn} + j\tilde{U}_{mn} \\
\tilde{Q}_{mn} - j\tilde{U}_{mn} & \tilde{I}_{mn} - \tilde{V}_{mn}
\end{pmatrix}.
\end{equation}
Thus, the source's brightness distribution of each Stokes parameter can be derived once $\G$ and $\D$ matrices are estimated and removed from the data. $\Par$ consists of purely geometric terms and can be determined straightforwardly.

Nearly all calibrators are resolved with VLBI, and their full polarization structures change rapidly (typically over several weeks or months; for example, \citealt{Jorstad2017, Lister2018, Weaver2022}). This means that $\bar\V$ are unknown and cannot be easily modeled using a simple assumption such as $\tilde{I}$, $\tilde{Q}$, $\tilde{U}$, and $\tilde{V}$ are proportional to each other\footnote{This assumption holds when calibrators are point sources, which is typical for connected interferometers.}. As a result, a standard calibration strategy for VLBI data has been to estimate $\G$ first using the parallel-hand visibilities and then $\D$ using the cross-hand visibilities after removing $\G$. It is because parallel-hand visibilities are sensitive to source's total intensity emission and antenna gains, but insensitive to linear polarization emission and D-terms, whereas cross-hand visibilities are the opposite. GPCAL follows this strategy as well \citep{Park2021a}. Some recently developed calibration/imaging pipelines, such as \texttt{polsolve} \citep{MartiVidal2021}, THEMIS \citep{Broderick2020}, DMC \citep{Pesce2021}, and Comrade \citep{Tiede2022}, utilize the full RIME for fitting.

The antenna field-rotation angles are usually corrected at an upstream calibration stage (before performing global fringe fitting). Thus, Equation~\ref{eq:rime} becomes:
\begin{equation}
    \V_{mn} = \Par^{-1}_m \G_m \D_m \Par_m \bar\V_{mn} \Par^H_n \D^H_n \G^H_n (\Par^H_n)^{-1},
    \label{eq:matmodel}
\end{equation}
which can be rewritten element-wise as \citep[e.g.,][]{Roberts1994}:
\begin{eqnarray}
\label{eq:model}
r^{RR}_{mn} &=& G^{R}_{m}G^{R*}_{n}[\mathscr{RR} + D^R_me^{2j\phi_m}\mathscr{LR} +\nonumber\\
&& D^{R*}_ne^{-2j\phi_n}\mathscr{RL} + D^R_{m}D^{R*}_{n}e^{2j(\phi_m-\phi_n)}\mathscr{LL}] \nonumber \\
r^{LL}_{mn} &=& G^{L}_{m}G^{L*}_{n}[\mathscr{LL} + D^L_me^{-2j\phi_m}\mathscr{RL} +\nonumber\\
&& D^{L*}_ne^{2j\phi_n}\mathscr{LR} + D^L_{m}D^{L*}_{n}e^{-2j(\phi_m-\phi_n)}\mathscr{RR}] \nonumber \\
r^{RL}_{mn} &=& G^{R}_{m}G^{L*}_{n}[\mathscr{RL} + D^R_me^{2j\phi_m}\mathscr{LL} +\nonumber\\
&& D^{L*}_ne^{2j\phi_n}\mathscr{RR} + D^R_{m}D^{L*}_{n}e^{2j(\phi_m+\phi_n)}\mathscr{LR}] \nonumber \\
r^{LR}_{mn} &=& G^{L}_{m}G^{R*}_{n}[\mathscr{LR} + D^L_me^{-2j\phi_m}\mathscr{RR} +\nonumber\\
&& D^{R*}_ne^{-2j\phi_n}\mathscr{LL} + D^L_{m}D^{R*}_{n}e^{-2j(\phi_m+\phi_n)}\mathscr{RL}].\nonumber\\
\end{eqnarray}

The last two equations and their approximations are used in GPCAL \citep{Park2021a} and LPCAL \citep{Leppanen1995}, respectively. In the original GPCAL pipeline, the D-terms are assumed to remain constant over frequency within the bandwidth and over time. We present new methods for calibrating frequency- (the present paper) and time-dependent polarimetric leakages (Paper II) that have been implemented in GPCAL.

\section{Calibration Procedure}
\label{sec:pipeline}

We present the method of correcting for frequency-dependent polarimetric leakages in VLBI data. The method assumes that antenna gains are unity, same as the original GPCAL pipeline, assuming that they are accurately constrained during the upstream calibration procedure and the imaging and self-calibration procedure. There is still a single phase offset ($e^{j\phi_{RL, \rm ref}}$) remaining in the cross-hand visibilities of all baselines, which originates from the instrumental phase offset between polarizations at the reference antenna.

The equations of our interest are:
\begin{eqnarray}
\label{eq:fitmodel}
r^{RL}_{mn}(\nu) &=& \left[\tilde{Q}_{mn}(\nu) + j\tilde{U}_{mn}(\nu)\right] + D^R_m (\nu) e^{2j\phi_m}r^{LL}_{mn, {\rm cal}} (\nu) \nonumber \\ &+& D^{L*}_n(\nu)e^{2j\phi_n}r^{RR}_{mn, {\rm cal}} (\nu) \nonumber \\ &+& D^R_{m}(\nu)D^{L*}_{n}(\nu)e^{2j(\phi_m+\phi_n)}\left[\tilde{Q}_{mn}(\nu) - j\tilde{U}_{mn}(\nu)\right] \nonumber \\
r^{LR}_{mn} (\nu) &=& \left[\tilde{Q}_{mn}(\nu) - j\tilde{U}_{mn}(\nu)\right] + D^L_m (\nu) e^{-2j\phi_m}r^{RR}_{mn, {\rm cal}} (\nu) \nonumber\\ &+& D^{R*}_n (\nu) e^{-2j\phi_n}r^{LL}_{mn, {\rm cal}} (\nu) \nonumber\\ &+& D^L_{m} (\nu) D^{R*}_{n} (\nu) e^{-2j(\phi_m+\phi_n)}\left[\tilde{Q}_{mn}(\nu) + j\tilde{U}_{mn}(\nu)\right]. \nonumber\\
\end{eqnarray}
$\mathscr{RL}$ and $\mathscr{LR}$ are replaced by Equation~\ref{eq:stokes} and $\mathscr{RR}$ and $\mathscr{LL}$ by the final calibrated parallel-hand visibilities $r^{RR}_{mn, {\rm cal}}$ and $r^{LL}_{mn, {\rm cal}}$, respectively. It is assumed that $e^{j\phi_{RL, {\rm ref}}}$ has already been corrected, which can be done by comparing the integrated electric vector position angles (EVPAs) of calibrators to known values. This can usually be accomplished by using sources with stable integrated EVPAs over time or from contemporaneous observations of single dishes or connected arrays. $\tilde{Q}_{mn}(\nu)$ and $\tilde{U}_{mn}(\nu)$ are obtained from the Fourier Transforms of the source's linear polarization images. In order to obtain these images, CLEAN is applied to the polarimetric leakage corrected data acquired by running the GPCAL pipeline on the frequency-averaged data. It is assumed that the linear polarization images are constant over frequency for each baseband channel (often referred to as ``IF"). We make the assumption that the fluctuations of polarimetric leakages across different frequencies are more significant than those of the source's intrinsic linear polarization emission. To obtain the values of the former, we consider the latter to be a constant and solve for it accordingly\footnote{In spite of this, $\tilde{Q}_{mn}(\nu)$ and $\tilde{U}_{mn}(\nu)$ are not exactly constant over frequency since the visibility at different $\nu$ has a different $(u,v)$, although the variation due to the $(u,v)$ coverage effect is generally expected to be small.}. Field-rotation angles are independent of the frequency.

The method fits Equation~\ref{eq:fitmodel} to the data for each channel to determine $D^R(\nu)$ and $D^L(\nu)$ using the Scipy curve\_fit package. As with the GPCAL pipeline, it is possible to fit the model to data from multiple calibrator sources simultaneously to enhance the fitting accuracy.

The method uses a multi-source UVFITS dataset (AIPS data catalog file) provided by users. To reduce calculation time and memory requirements, it bins the data in frequency, if specified by users. This approach is usually recommended because recent VLBI data often contain a large number of channels, while the D-terms are unlikely to vary significantly across adjacent channels. Determining the optimal bin size may be data-dependent and therefore, it is recommended that users determine this parameter themselves. If the phase offset ($e^{j\phi_{RL, {\rm ref}}}$) still exists in the provided data, the method can remove the offset using the offset value provided by users\footnote{As explained above, the offset value can be derived by performing the EVPA calibration of the frequency-averaged data.}, which may vary depending on the IF used. The multi-source UVFITS data is usually not self-calibrated. The method performs amplitude and phase self-calibration with the task CALIB in AIPS, if requested, using the total intensity CLEAN models (saved in Difmap fits image files) provided by users. 

The method runs the GPCAL pipeline to derive the D-terms for each channel. It then calibrates the multi-source UVFITS data using the derived D-term spectra to obtain $\bar{V}$ in Equation~\ref{eq:rime}. This results in the final output being a multi-source UVFITS dataset saved in an AIPS catalog after correction of the frequency-dependent D-terms.

The delay offset between polarizations at the reference antenna is corrected during data pre-processing in AIPS after global fringe fitting. It should be noted, however, that the frequency-dependent instrumental polarization may prevent accurate delay offset corrections. Thus, it is recommended to perform additional delay offset corrections using the output data after the frequency-dependent D-terms have been removed. It is recommended that users perform this additional correction after checking the output file, as it is not included in the calibration procedure. The method is implemented in GPCAL, which is publicly available at \url{https://github.com/jhparkastro/gpcal.}

\section{Validation using Synthetic Data}
\label{sec:synthetic}

We validate our method by employing synthetic data generated from actual VLBA data. Our objective is to evaluate the effectiveness of our method on synthetic data that closely resembles real data in terms of properties. To achieve this, we acquired the ground-truth models of total intensity and linear polarization for the sources, as well as the noise characteristics and frequency-dependent leakages from real VLBA 15 GHz data. These real data components served as the basis for generating the synthetic data. The data we selected is part of the Monitoring of Jets in Active Galactic Nuclei with VLBA Experiments\footnote{\url{https://www.physics.purdue.edu/MOJAVE/}} (MOJAVE; \citealt{Lister2018}) data sets, which have been monitoring a number of AGN jets with the VLBA at 15 GHz for decades. The data were collected from the observation of 25 sources on 2020 June 14 (project code: BL273AK). The data were recorded in both RCP and LCP, with two-bit quantization in four IFs, using the digital downconverter (DDC) observing system, at a recording rate of 4 Gbps, yielding a total bandwidth of 128 MHz for each IF and polarization.

We performed a standard data postcorrelation process with AIPS according to \cite{Park2021b} and hybrid imaging with CLEAN and self-calibration in Difmap. We ran the GPCAL pipeline on the self-calibrated data. We used several calibrators with linear polarization structures dominated by cores for the initial D-term estimation using the similarity approximation. Further instrumental polarization self-calibration was performed with 10 iterations using several calibrators with high signal-to-noise ratios in the cross-hand visibilities. The phase offset between polarizations at the reference antenna was corrected by comparing the integrated EVPAs of several calibrators using the D-term corrected data provided by GPCAL with the data already calibrated in the monitoring program's public database. We produced Stokes $Q$ and $U$ images of the sources with CLEAN in Difmap.

\begin{figure*}[t!]
\centering
\includegraphics[width = \textwidth]{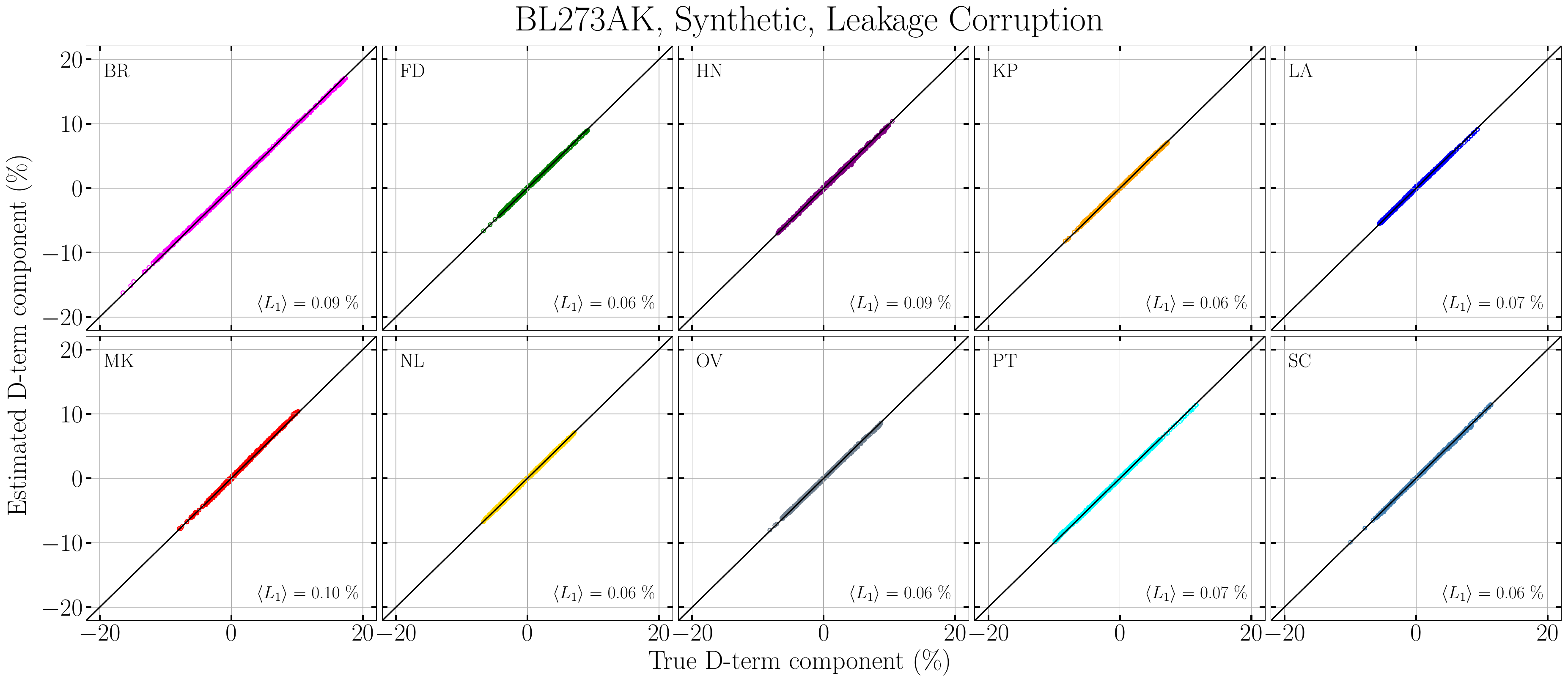}
\caption{Comparison of the ground-truth D-term components (real and imaginary parts) assumed in the synthetic data sets shown on x-axis with the reconstructed D-term components derived by the method shown on y-axis for each frequency channel. Each station's result is presented using distinct colors. For each antenna, the correlation between the actual and estimated D-term components is shown in units of percentage. The norm $L_1 \equiv |D_{\rm Truth} - D_{\rm Recon}|$ is averaged over real and imaginary components of the D-terms for RCP and LCP for all antennas, and is denoted in each panel. \label{fig:dsyn1to1}}
\end{figure*}

\begin{figure*}[t!]
\centering
\includegraphics[width = \textwidth]{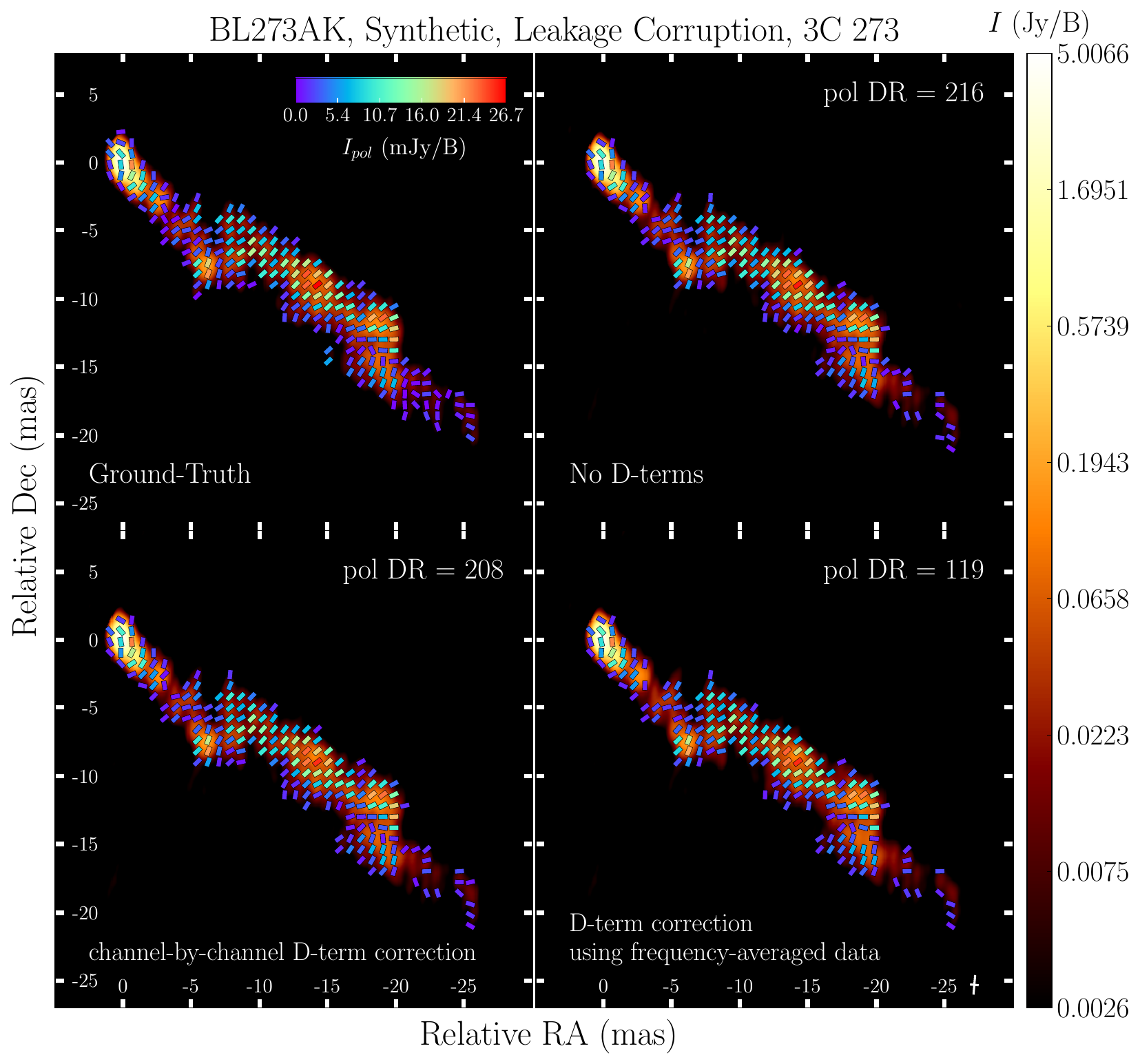}
\caption{Linear polarization images of 3C 273 from the synthetic data set generated based on the real VLBA 15 GHz data. Colored ticks indicate EVPAs, with colors corresponding to linearly polarized intensities. We have made Ricean de-biasing corrections (i.e., $P_{\rm corr} = P_{\rm obs}\sqrt{1 - (\sigma_P / P_{\rm obs})}$) to the images \citep{WK1974}, where $P_{\rm corr}$ and $P_{\rm obs}$ denote the corrected and original linearly polarized intensity, respectively. The white crosses in the bottom right corner represent the shape of the synthesized beam. The top left image is the ground-truth image used for synthetic data generation. The top right image is obtained from the data without being corrupted by polarimetric leakage (``No D-Terms"). Hence, this image comprises solely of two potential sources of error, namely thermal noise within the data and inaccuracies caused by the utilization of the CLEAN algorithm. Consequently, this image could serve as a benchmark reference for the purpose of scrutinizing the influence of polarimetric leakages within the data. In the bottom left image, we have applied our method of frequency-dependent instrumental polarization calibration to data where frequency-dependent leakage corruption has been introduced (``channel-by-channel D-term correction"). Using the same data, the bottom right image was produced by averaging the data over frequency for each IF and then performing instrumental polarization calibration based on the averaged data (``D-term correction using frequency-averaged data"). A comparison of the bottom left and right images can be used to evaluate the effect of frequency-dependent leakage calibration. With the exception of the ground-truth image, which is noise-free, each panel exhibits the dynamic range of the linear polarization images. \label{fig:dsynimage}}
\end{figure*}

\begin{figure*}[t!]
\centering
\includegraphics[width = \textwidth]{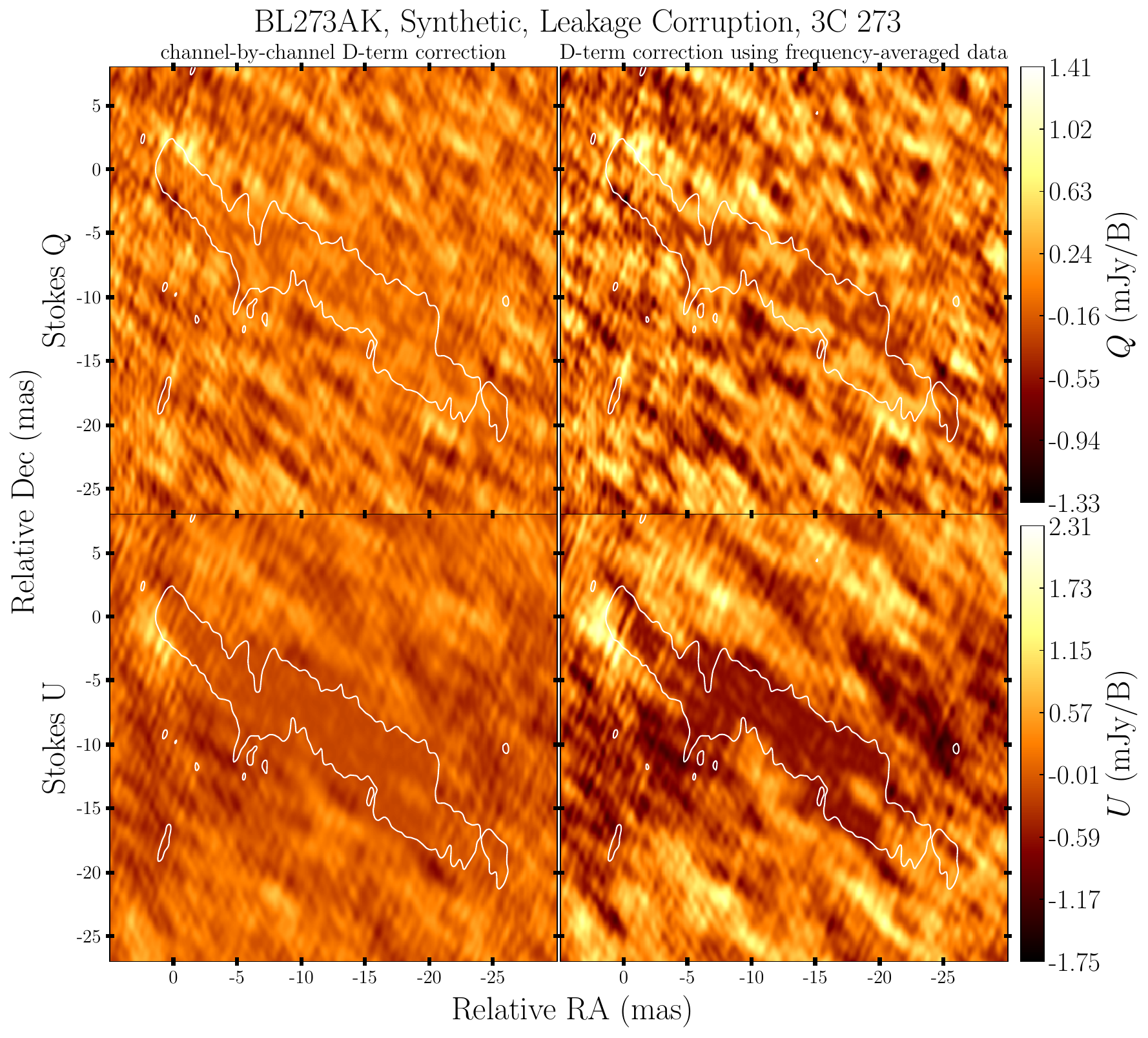}
\caption{Dirty images of the residual visibilities of synthetic data for 3C 273, which represent the visibilities obtained after subtracting the CLEAN model visibilities for Stokes $Q$ (upper) and $U$ (lower). White contours represent the contour of the Stokes I image, with the contour level set at three times the off-source image rms noise, delineating the source region. Images on the left, obtained with frequency-dependent D-term correction, exhibit significantly reduced noise levels compared to those on the right, which use frequency-averaged data for D-term correction. \label{fig:dsyndirty}}
\end{figure*}

\begin{figure*}[t!]
\centering
\includegraphics[width = 0.49\textwidth]{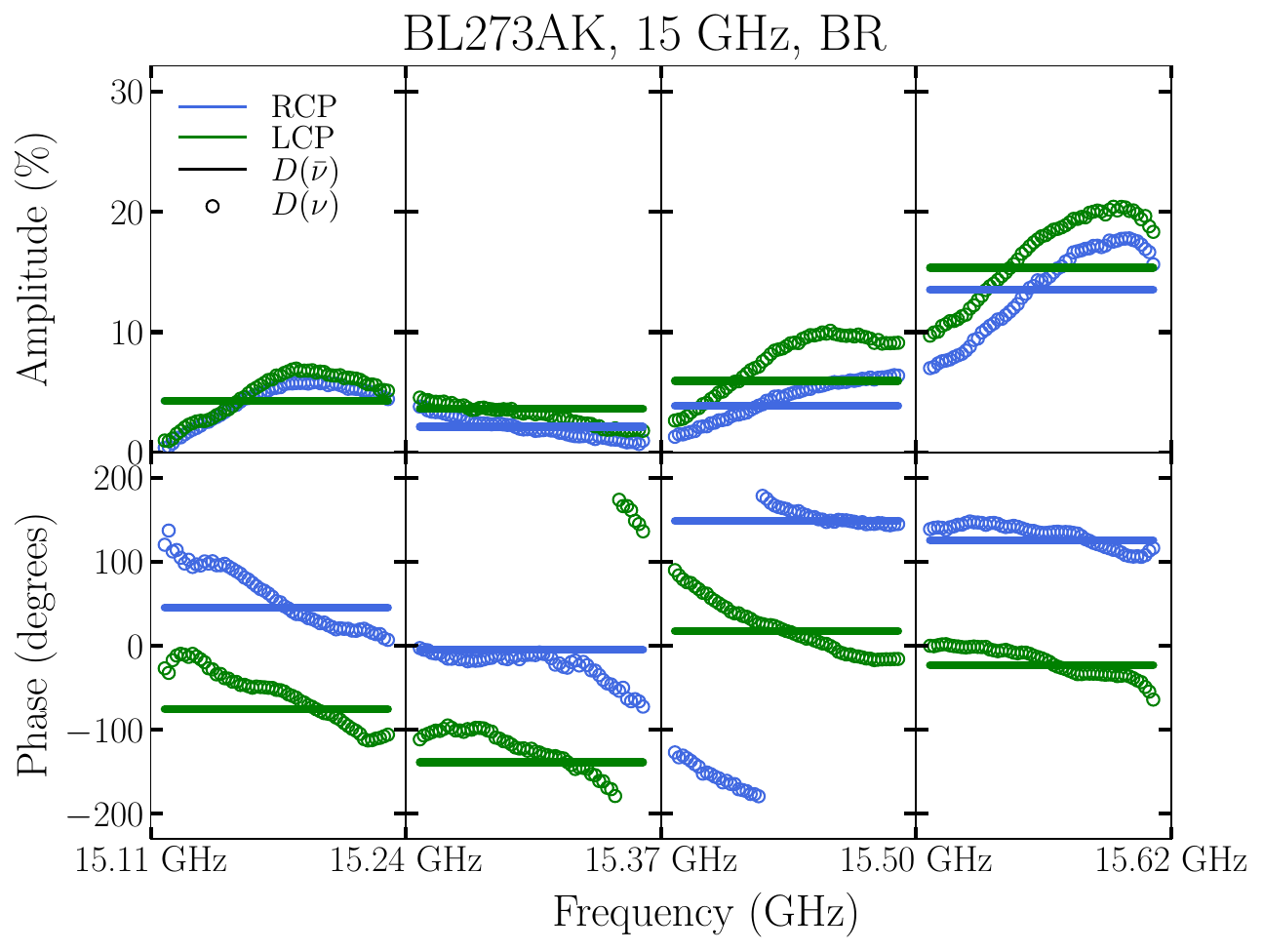}
\includegraphics[width = 0.49\textwidth]{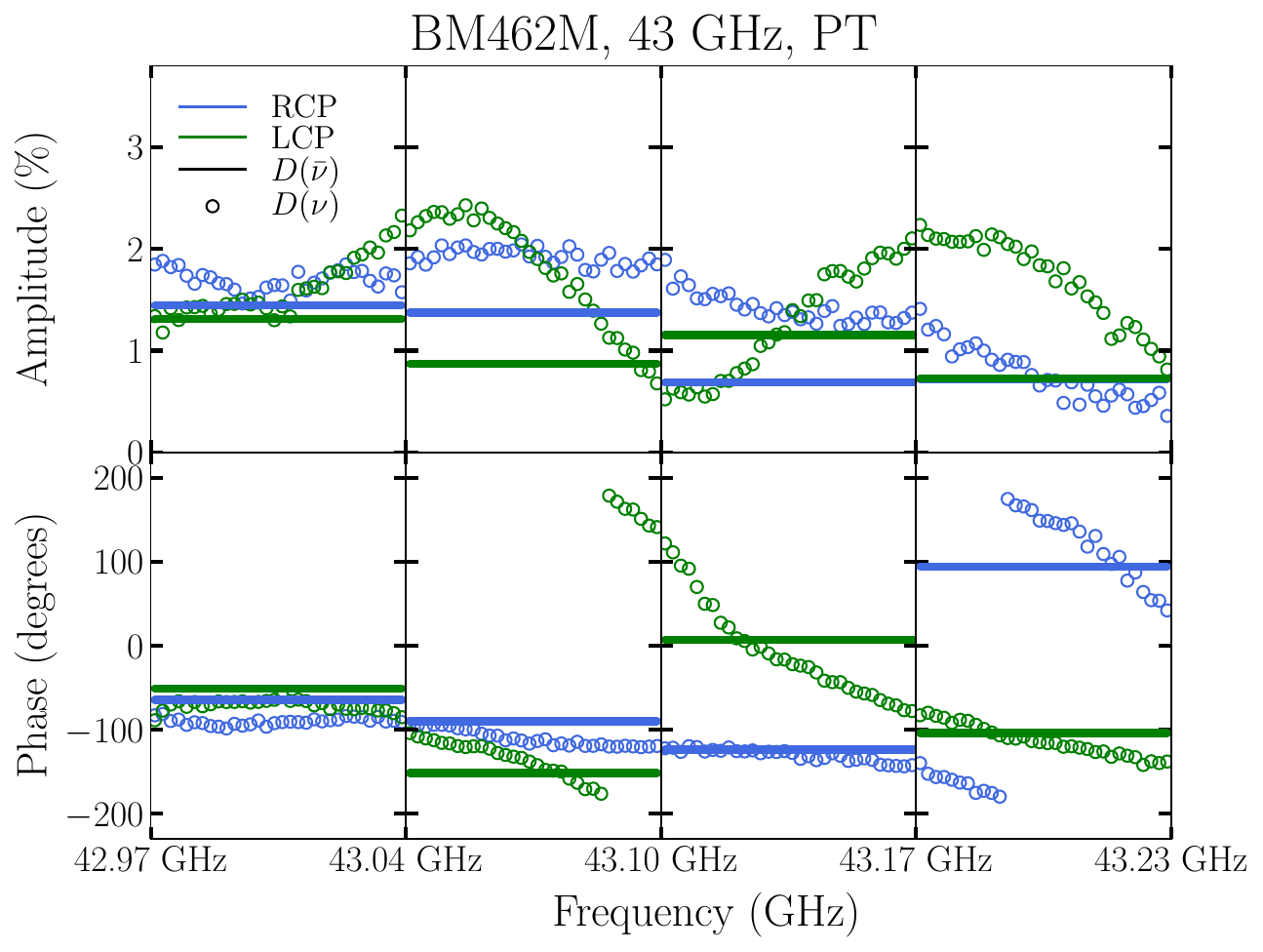}
\caption{D-terms for RCP (blue) and LCP (green) for BL273AK (left, 15 GHz) and BM462M (right, 43 GHz) data. The D-terms of specific stations, such as BR for BL273AK and PT for BM462M, are presented as examples. The open circles indicate the D-term spectra derived from GPCAL's frequency-dependent leakage calibration ($D(\nu)$), while the solid lines indicate D-terms obtained from the frequency-averaged data set ($D(\bar{\nu})$). We note that in the right figure, the green and blue solid lines representing the amplitudes of the 4th IF overlap, making them difficult to distinguish from each other. The D-term spectra exhibit smooth and continuous variation over frequency within and across IFs. The $D(\bar{\nu})$ is a good approximation of the $D(\nu)$ for both data sets. It is worth noting a noticeable discrepancy between $D(\bar{\nu})$ and $D(\nu)$ in the right figure. However, the numerical difference is relatively small, and it is important to emphasize that $D(\bar{\nu})$ is not necessarily an average of $D(\nu)$, as discussed in Appendix~\ref{appendix:linear}.. \label{fig:dspec}}
\end{figure*}

We generated synthetic data for three sources, 3C 273, 3C 279, and OJ 287, using the synthetic data generation function implemented in GPCAL, as described by \cite{Park2021c}. In brief, the function generates a synthetic data set based on Equation~\ref{eq:matmodel}, using the $I$, $Q$, and $U$ CLEAN models as ground-truth source structures (i.e., their Fourier Transforms are $\tilde{I}$, $\tilde{Q}$, $\tilde{U}$ in Equation~\ref{eq:stokes}), and assuming $\tilde{V}=0$. The multi-channel real VLBA 15 GHz data are replaced by the source model visibility corresponding to each $(u,v)$, and we added thermal noise based on the uncertainties of the real data for each channel. We introduce frequency-dependent D-terms, which were derived from the real data using GPCAL (Section~\ref{sec:application}). For example, the frequency-dependent leakages of the VLBA BR station shown in the left panel of Figure~\ref{fig:dspec} are introduced to the synthetic data set of the same station using Equation~\ref{eq:matmodel}. Since the purpose of the synthetic data test is to validate our method of correcting for frequency-dependent leakages, we assumed unity antenna gains, i.e., $\G = \I$ (However, refer to the discussion below and Appendix~\ref{appendix:synthetic} for the analysis of synthetic data, incorporating antenna gain distortions.).

From the synthetic data set, we derived frequency-dependent D-terms with GPCAL. Firstly, the synthetic data set was averaged over frequency for each IF. The total intensity CLEAN models are reconstructed using Difmap for each source from this data set. Secondly, we corrected antenna leakages by running the GPCAL pipeline on the frequency-averaged data, from which we obtained Stokes $Q$ and $U$ CLEAN images. For this leakage correction, an initial D-term estimation is performed with OJ 287 based on the similarity approximation, followed by instrumental polarization self-calibration with 10 iterations using all three sources. Thirdly, frequency-dependent leakages were derived with the method from the frequency-unaveraged data set and using the $Q$ and $U$ CLEAN images. For this estimation, all three sources were used simultaneously. Lastly, we averaged the frequency-dependent leakage corrected data set over frequency and determined Stokes $Q$ and $U$ images based on that data set.

In Figure~\ref{fig:dsyn1to1}, we compare the components of the ground-truth D-terms (real and imaginary parts) with those of the reconstructed D-terms. Each panel indicates an average of the $L_1 \equiv |D_{\rm Truth} - D_{\rm Recon}|$ norm. The method was able to reproduce ground-truth D-terms with an accuracy of an $\langle L_1\rangle$ norm of $\lesssim0.1\%$ for all VLBA stations. 

Figure~\ref{fig:dsynimage} showcases the impact of frequency-dependent leakage correction on the resulting linear polarization images of 3C 273, as compared to the ground-truth linear polarization image employed in the generation of synthetic data (top left, labeled as ``Ground-Truth"). Aside from the linear polarization image obtained after correcting for frequency-dependent D-terms (bottom left, ``channel-by-channel D-term correction"), two reference polarization images were produced. The first is an image that has been reconstructed from a synthetic data set generated in accordance with the above description, but without taking into account leakage (top right, ``No D-terms"). There are only two sources of error in this image: thermal noise in the data and errors introduced by the CLEAN algorithm. As a result, this image could function as a standard point of reference for evaluating the impact of polarimetric leakages present in the data. Another image is reconstructed using the synthetic data set, however, instead of a frequency-dependent leakage correction, the leakages are corrected using the data set averaged over frequency for each IF (bottom right, ``D-term correction using frequency-averaged data"). It is through this method that LPCAL and the original GPCAL pipeline would perform leakage calibration. This image can then be used as a reference to test the effect of frequency-dependent leakage calibration.

In each panel, we note the dynamic range (DR) of the linear polarization images, defined as $P_{\rm peak} / \sigma_P$, where $P_{\rm peak}$ represents the peak polarization intensity and $\sigma_P \equiv (\sigma_Q + \sigma_U)/2$ represents the average of the rms noise for the off-source regions in the Stokes $Q$ and $U$ images \citep{Hovatta2012}. The DR of the image obtained with frequency-dependent D-term correction (bottom left) is slightly lower than that of the top right image, while the DR of the image obtained with leakage correction using the frequency-averaged data (bottom right) is about half that of the top right image. 

In Figure~\ref{fig:dsyndirty}, we present the dirty images of the residual visibilities, which are the visibilities after subtracting the CLEAN model visibilities for Stokes $Q$ and $U$. These images showcase the noise levels within the residual visibilities, which can serve as an indicator of the accuracy of the leakage calibration. It is evident that the application of frequency-dependent leakage correction has resulted in a significant reduction in image noise compared to the image obtained with leakage correction using frequency-averaged data. Based on this difference, it can be concluded that frequency-dependent leakage correction with \jp{the method} improves the quality of linear polarization images significantly when antenna gains are perfectly calibrated.

\section{Application to Real Data}
\label{sec:application}

\subsection{Data \& Analysis}

In this section, we present the results of applying the frequency-dependent leakage calibration method to two real VLBA data sets. The first is the same VLBA 15 GHz data set that was presented in Section~\ref{sec:synthetic}. The second is part of the VLBA-BU-BLAZAR monitoring program\footnote{\url{https://www.bu.edu/blazars/VLBAproject.html}}, which has monitored many $\gamma$-ray bright AGN jets with the VLBA at 43 GHz for decades. We used the data from the observation of 33 sources on 2018 December 8 (project code: BM462M). The data were recorded in both RCP and LCP with two-bit quantization in four IFs, using the digital downconverter (DDC) observing system, at a recording rate of 2 Gbps, yielding a total bandwidth of 64 MHz for each IF and polarization.

We first obtained source's total intensity and linear polarization images from the frequency-averaged data set as described in Section~\ref{sec:synthetic}. Based on these images, frequency-dependent D-terms were derived using the method following the same procedure described in Section~\ref{sec:synthetic}. We used multiple bright calibrators simultaneously to derive the D-terms (OJ 287, 3C 273, and 3C 279 for the 15 GHz data and OJ 287, 3C 273, 3C 279, 3C 84, and 3C 454.3 for the 43 GHz data). After correcting for frequency-dependent D-terms, we averaged the data set over frequency and ran the GPCAL pipeline. This is because we derived frequency-dependent D-terms using the data for each channel, which have smaller S/Ns than the frequency-averaged data, D-terms with small amplitudes may remain in the data set.

\begin{figure*}[t!]
\centering
\includegraphics[width = 1.0\textwidth]{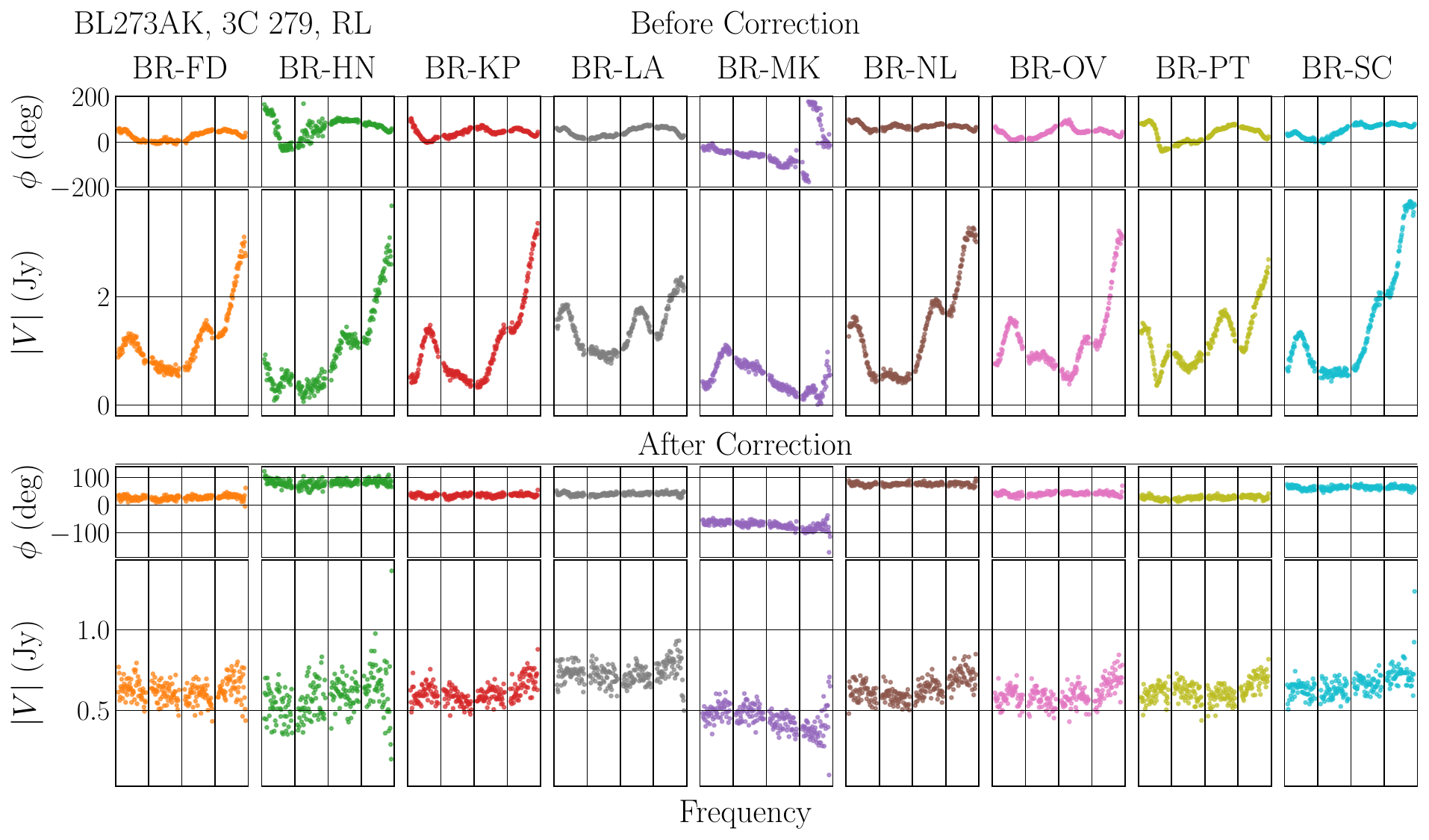}
\includegraphics[width = 1.0\textwidth]{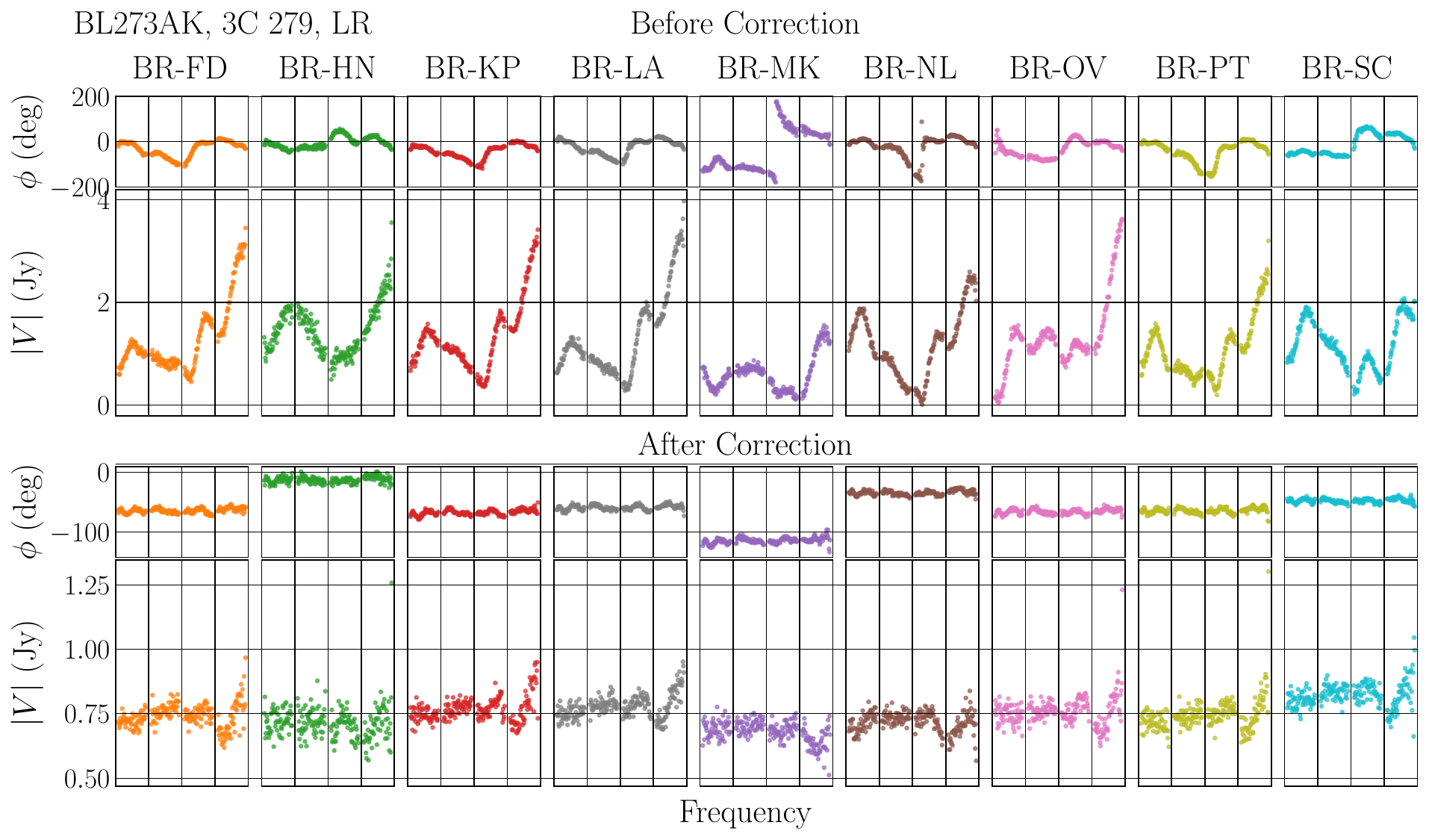}
\caption{Cross power spectra (amplitude; $|V|$, phase; $\phi$) of the cross-hand visibilities (RL; upper, LR; lower) of 3C 279 obtained from the BL273AK data. The baselines associated with BR station averaged over one scan are shown. The upper and lower spectra are presented before and after correcting for frequency-dependent polarimetric leakage, respectively. \label{fig:possm_bl273ak}}
\end{figure*}

\begin{figure*}[t!]
\centering
\includegraphics[width = 1.0\textwidth]{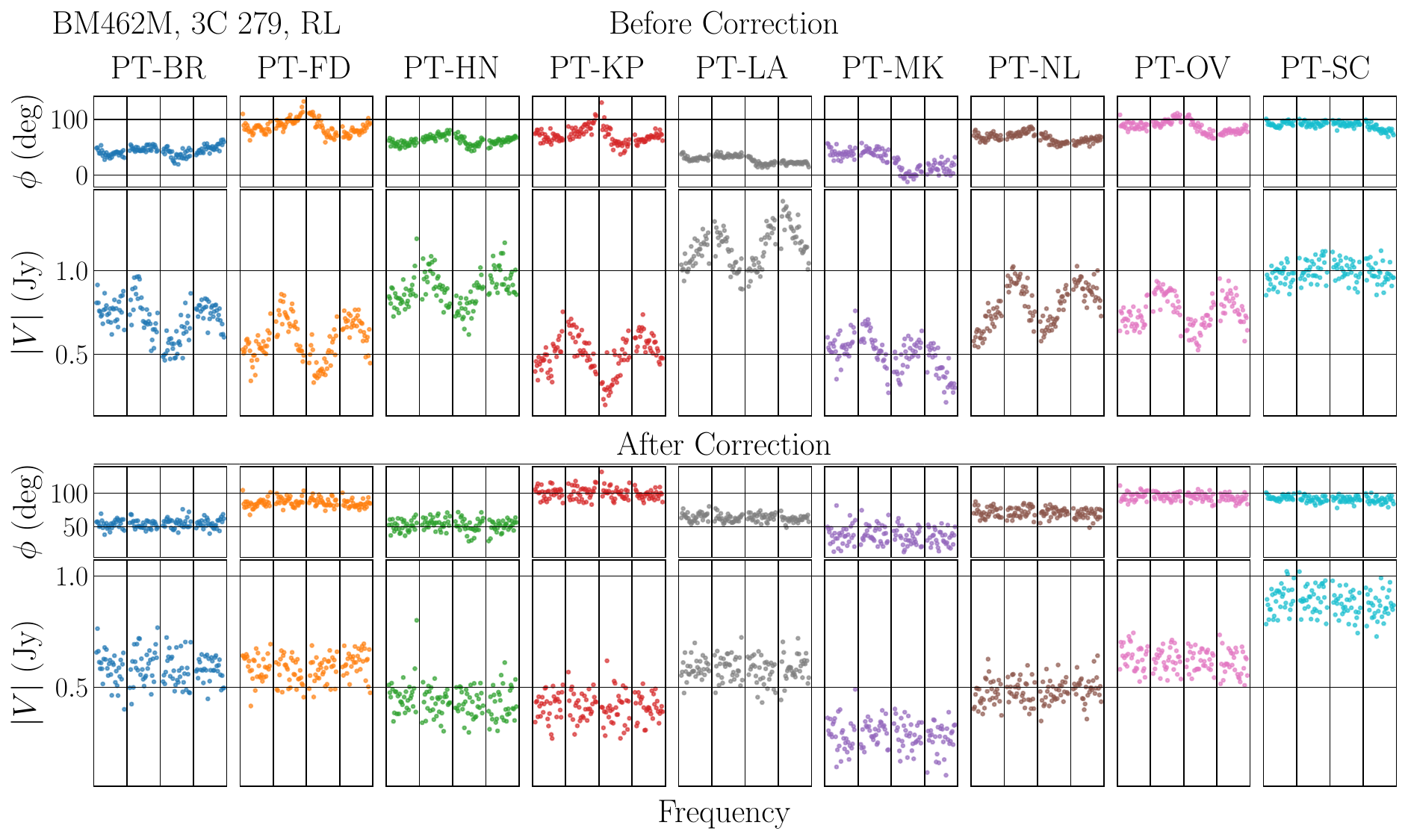}
\includegraphics[width = 1.0\textwidth]{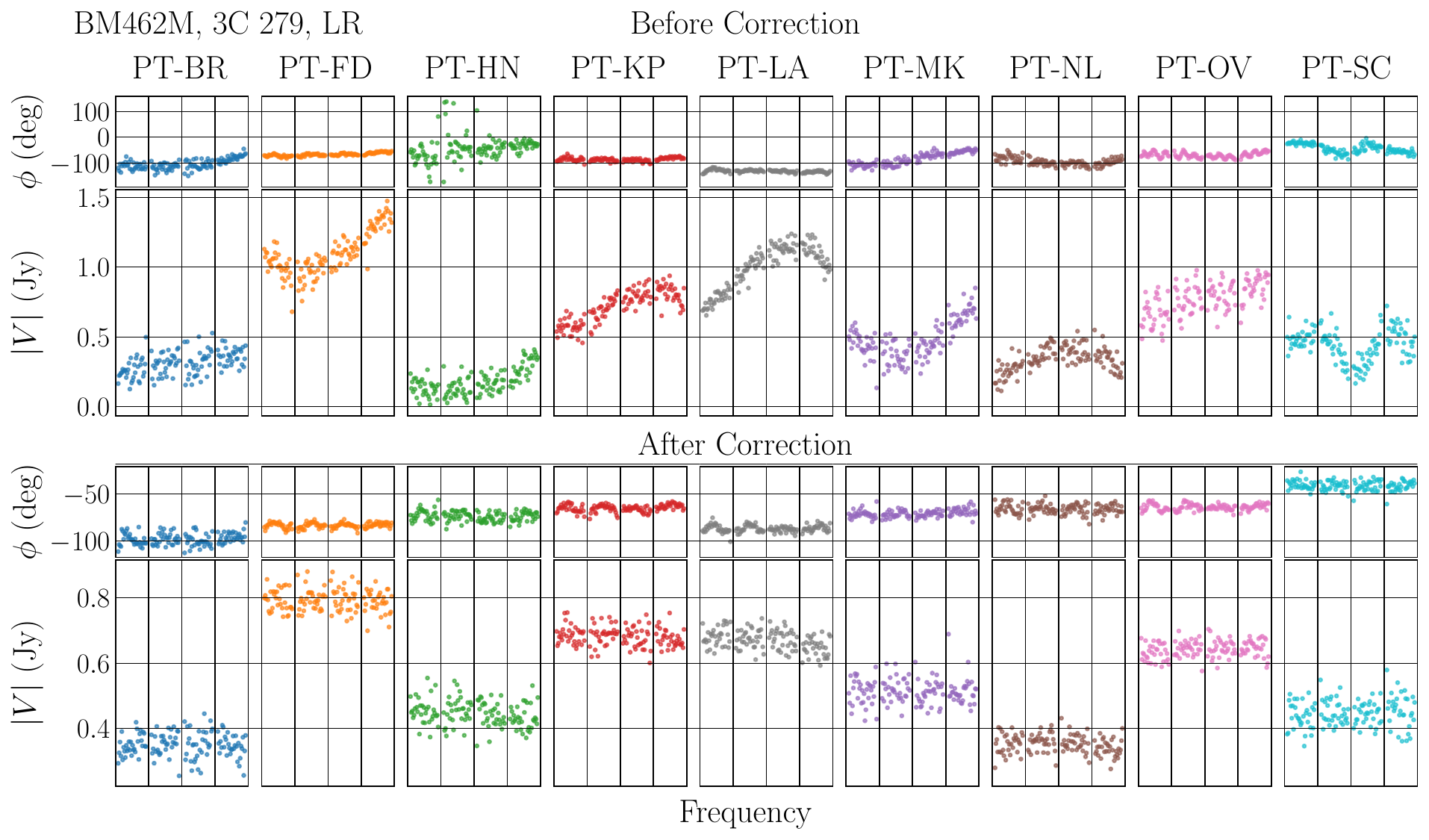}
\caption{Same as Figure~\ref{fig:possm_bl273ak} but for PT baselines of the BM462M data observed at 43 GHz. \label{fig:possm_bm462m}}
\end{figure*}

\begin{figure*}[t!]
\centering
\includegraphics[width = 1.0\textwidth]{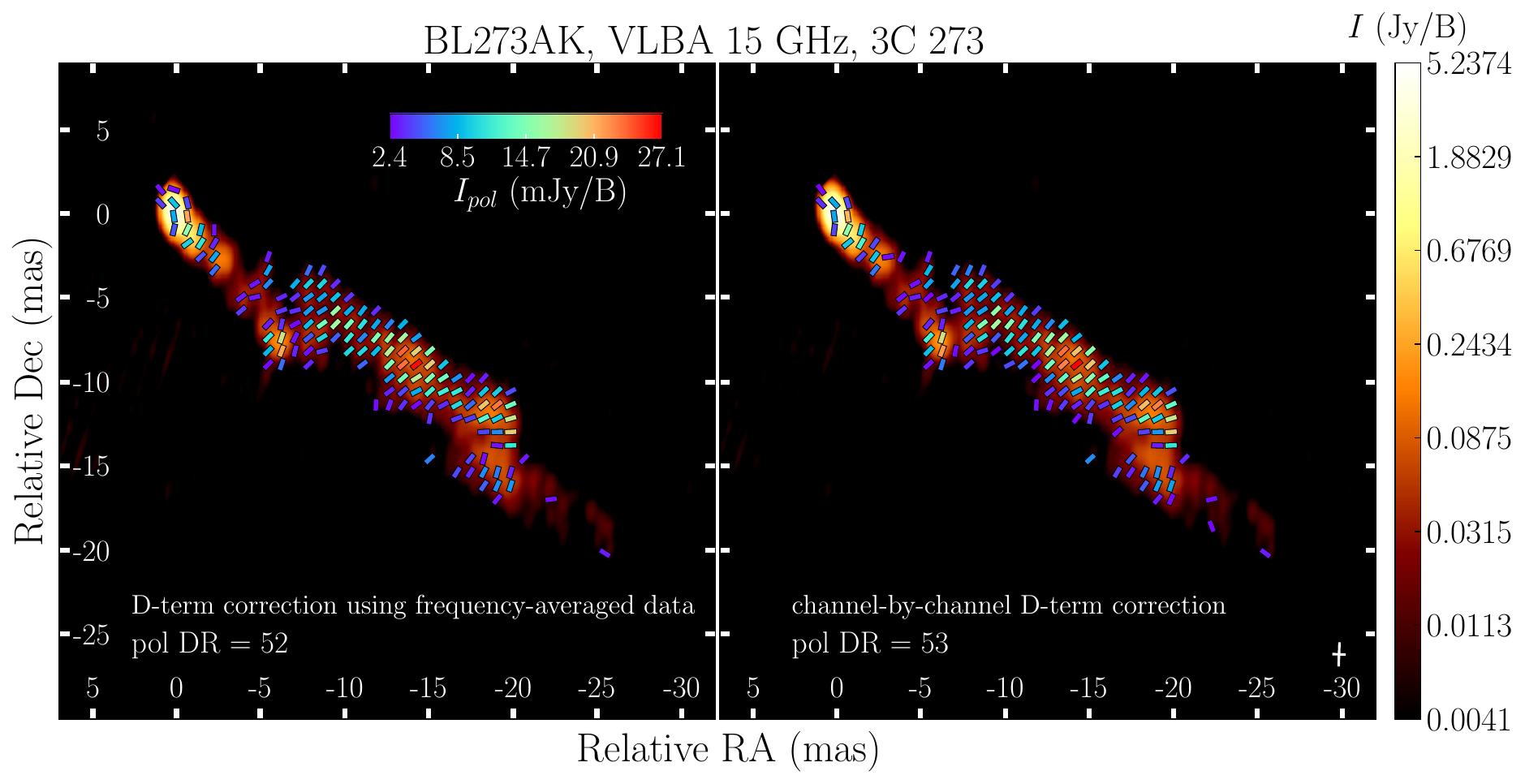}
\includegraphics[width = 1.0\textwidth]{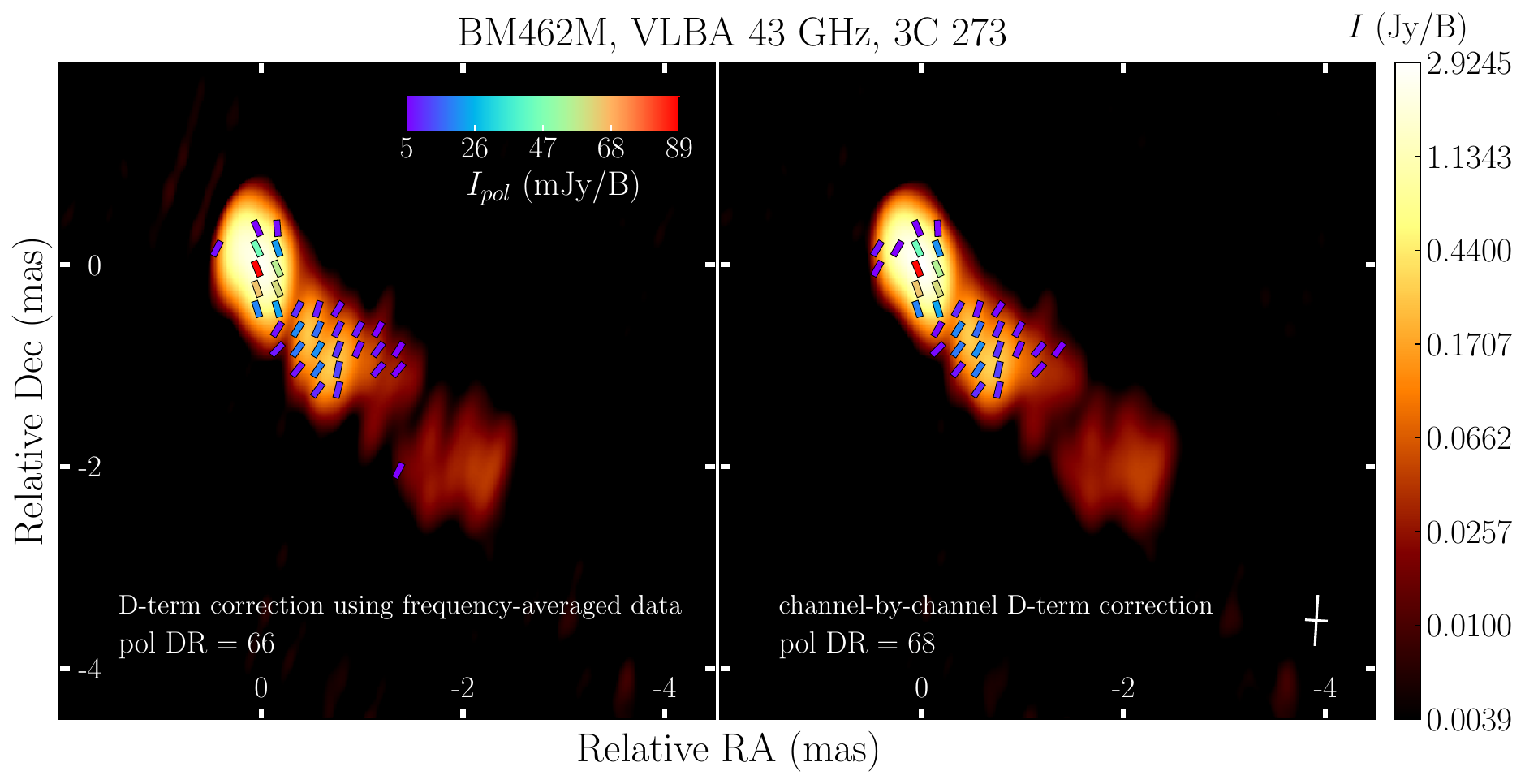}
\caption{Linear polarization images of 3C 273 from the BL273AK data (upper, 15 GHz) and BM462M data (lower, 43 GHz). Colored ticks indicate EVPAs, with colors corresponding to linearly polarized intensities. The white crosses in the bottom right corner of the right panel represent the shape of the synthesized beam. The images obtained by using the GPCAL pipeline using the data averaged over frequency within eacn IF are presented in the left panels. The images obtained by correcting frequency-dependent polarimetric leakages using the method presented in this paper are shown in the right panels. The polarization image DR is noted in each panel. \label{fig:map}}
\end{figure*}

\subsection{Results}

In Figure~\ref{fig:dspec}, we present the derived D-term spectra ($D(\nu)$) for BR and PT stations for the BL273AK and BM462M data sets, respectively. The D-terms derived from the frequency-averaged data using GPCAL are overplotted ($D(\bar{\nu})$; solid lines). In both data sets, the D-term spectra display smooth variation over frequency across different IFs, indicating that these spectra accurately reflect the characteristics of the instruments. As expected, $D(\bar{\nu})$ follows $D(\nu)$ closely, but only approximates the rapidly varying spectrum. 

Figures~\ref{fig:possm_bl273ak} and~\ref{fig:possm_bm462m} show the cross-hand visibilities of 3C 279 averaged over one scan (with a scan length of $\sim1.5$ and $\sim4.5$ minutes for BL273AK and BM462M data, respectively) for both data sets before and after correction for frequency-dependent polarimetric leakages. We present data on the baselines associated with specific stations (BR station for BL273AK and PT station for BM462M) as an example. Cross-hand visibility amplitudes of many BR baselines in the BL273AK data exhibit a similar pattern, which is a rapid increase in amplitude with frequency for IF 4\footnote{It is not surprising to observe the rapid increase for IF 4, which has a frequency coverage of 15.496 -- 15.624 GHz. The nominal frequency range of the VLBA 2\,cm band is 12.0 -- 15.4 GHz according to the VLBA Observational Status Summary (\url{https://science.nrao.edu/facilities/vlba/docs/manuals/oss/bands-perf}). The IF is near the band edge, where the polarizer characteristic can become poor.}. It is noteworthy that this behavior is very similar to the trend of the derived D-term spectra for BR station (the left panel of Figure~\ref{fig:dspec}). This indicates that the frequency-dependent polarimetric leakage of this station is responsible for the pattern in the cross-hand visibilities. The behavior disappears when the D-terms are removed from the data (lower panel). Similar results were obtained for the BM462M data as well (Figure~\ref{fig:possm_bm462m}). As an example, after the correction, the zigzag pattern in the RL baselines associated with PT station has disappeared. In conclusion, these results demonstrate that the frequency-dependent calibration method implemented in GPCAL is effective.

\subsection{Evaluation}
\label{sec:evaluation}

In this subsection, we evaluate the effects of frequency-dependent instrumental polarization calibration on linear polarization images. In Figure~\ref{fig:map}, we present the linear polarization image of the bright quasar 3C 273, which has a complex linear polarization structure, after correcting for frequency-dependent D-terms. It is compared with the image obtained from frequency-averaged data. It was found that the linear polarization images obtained with and without frequency-dependent leakage correction were almost identical. The results showed that the polarization image dynamic range slightly improved with the correction. The results for other sources were similar. 

According to our results, frequency-dependent instrumental polarization calibration can slightly improve the quality of resulting linear polarization images in some cases, however, the improvement is generally not very significant. Our hypothesis is that this is because the primary contribution of polarimetric leakages to the cross-hand visibilities is linear, and the distortion caused by averaging the data over frequency before correcting for frequency-dependent leakages is generally small (see Appendix~\ref{appendix:linear} for more details). Also, other systematic errors in the data, such as residual antenna gain errors, may dominate the errors caused by averaging data over frequency before correction for frequency-dependent leakage. In Appendix~\ref{appendix:synthetic}, we test this claim by using synthetic data sets that include both antenna gain and frequency-dependent leakage corruptions. In fact, we found that the quality of linear polarization images does not improve much with frequency-dependent instrumental polarization calibration, which is similar to what we found in real data sets. 

Generally, if D-term amplitudes are large and significantly vary over frequency within an IF, the distortion in cross-hand visibility caused by averaging data over frequency before correcting for frequency-dependent leakages (see Equation~\ref{eq:fitmodel}) will increase and degrade the quality of linear polarization images. Therefore, our method of frequency-dependent instrumental polarization calibration implemented in GPCAL will be particularly useful for calibrating such data. There has been a substantial increase in the bandwidth of modern radio interferometers in recent years. The calibration of frequency-dependent instrumental polarization becomes more critical for future VLBI data, where the D-terms are expected to vary significantly over a wide bandwidth. Connected interferometers, such as the Very Large Array and the Atacama Large Millimeter/submillimeter Array, have already observed this phenomenon \citep[e.g.][]{Nagai2016, Bower2018, Jagannathan2017, Jagannathan2018}.

\section{Conclusions}
\label{sec:conclusion}

Compared to the past, modern VLBI arrays offer a much wider bandwidth. The large bandwidth of the data allows for a significant reduction in thermal noise. However, systematic errors in the data do not decrease as bandwidth increases, and this can pose a significant limitation in terms of image quality. In the case of linear polarization images, this effect is more severe, since conventional calibration methods are not able to easily remove the systematic errors. Some of the systematic errors are polarimetric leakages varying over frequency within the bandwidth and over time. Leakages usually become larger at frequencies far from the nominal frequency in the band, as the instruments are designed to have small leakages at the nominal frequency. Also, polarization leakage is direction-dependent effect (see, e.g., lecture 3 in \citealt{Taylor1999}; \citealt{Thum2008, Smirnov2011}). Accordingly, D-terms can vary depending on the antenna pointing accuracy, which varies from scan to scan. Most existing calibration tools, however, assume that leakages are constant over time and frequency.

We present in this series of papers new calibration methods for frequency- (this paper) and time-dependent (Paper II) polarimetric leakages in VLBI data, which have been implemented in GPCAL. The method presented in this paper determines leakages for each channel based on the Stokes $Q$ and $U$ calibrator models, which are derived from the frequency-averaged data set. As in the original GPCAL pipeline, data from multiple calibrator sources can be used simultaneously for fitting to improve accuracy.

We verified the method using synthetic data generated based on real VLBA 15 GHz data. The method was able to reconstruct the ground-truth D-terms with an accuracy of $\langle L_1 \rangle \lesssim 0.1\%$. The linear polarization images are significantly improved with the frequency-dependent leakage calibration using the method in the case of no antenna gain corruption assumed in the synthetic data.

We applied the method to two sets of real data obtained with the VLBA at 15 and 43 GHz. The D-term spectra were derived using several bright and moderately to highly polarized calibrators. There is continuous and smooth variation in the spectra across all bandwidths, which indicates that the spectra are an accurate representation of the characteristics of the VLBA's instruments. The D-terms obtained from the data averaged over frequency within each IF are a good approximation of the D-term spectra. After correction of frequency-dependent D-terms, substantial variations in the cross-hand visibility spectra over frequency disappear. This result indicates that the variation is caused by frequency-dependent polarimetric leakages, rather than by intrinsic signals of the source. This result also indicates that the method is effective in removing the frequency-dependent leakages from real VLBI data.

We evaluated the effects of frequency-dependent instrumental polarization calibration on linear polarization images. There were no significant differences between the images obtained after correcting for frequency-dependent D-terms and those obtained from frequency-averaged data, although the dynamic range of the linear polarization images with frequency-dependent leakage corrections were slightly higher.

Based on these results, the calibration of D-terms based on frequency-averaged data appears to be a reasonable approximation for the particular VLBA data sets analyzed in this paper. The reason for this is that the primary contribution of D-terms to the cross-hand visibilities is linear. When the cross-hand visibilities are averaged over frequency before correcting frequency-dependent D-terms, there are second-order terms that can cause distortion. The distortion is expected to be small if the D-term amplitudes of frequency channels are small and do not vary significantly over frequency within each IF. Furthermore, other systematic errors in the data, such as residual antenna gain errors, may dominate the errors caused by averaging data over frequency before correcting for frequency-dependent leakage. This appears to be the case for the VLBA data sets presented in this paper. Nevertheless, we observe some improvement in the quality of polarization images as a result of calibration of frequency-dependent D-terms, indicating that users should consider calibration of frequency-dependent D-terms to increase calibration precision. Additionally, it will contribute significantly to the calibration of future VLBI data, for which significant variations in D-terms are expected over frequency within large bandwidths.

\acknowledgments

We express our gratitude to the anonymous referee for conducting a comprehensive review of our manuscript, which greatly enhanced the quality of the paper. J.P. acknowledges financial support through the EACOA Fellowship awarded by the East Asia Core Observatories Association, which consists of the Academia Sinica Institute of Astronomy and Astrophysics, the National Astronomical Observatory of Japan, Center for Astronomical Mega-Science, Chinese Academy of Sciences, and the Korea Astronomy and Space Science Institute. This work is supported by the Ministry of Science and Technology of Taiwan grant MOST 109-2112-M-001-025 and 108-2112-M-001-051 (K.A). The VLBA is an instrument of the National Radio Astronomy Observatory. The National Radio Astronomy Observatory is a facility of the National Science Foundation operated by Associated Universities, Inc. 

\facilities{VLBA (NRAO)}

\software{AIPS \citep{Greisen2003}, Difmap \citep{Shepherd1997}, GPCAL \citep{Park2021a}, ParselTongue \citep{Kettenis2006}, Scipy \citep{Scipy2020}}

\clearpage
\appendix

\section{Distortion in cross-hand visibilities caused by frequency-dependent polarimetric leakages}
\label{appendix:linear}

The original GPCAL pipeline uses the cross-hand visibilities averaged over frequency within each IF. This is equivalent to take average of the left and right hand sides of Equation~\ref{eq:fitmodel}:
\begin{eqnarray}
\label{eq:avgmodel1}
\langle r^{RL}_{mn}(\nu) \rangle_\nu &=& \langle\tilde{Q}_{mn}(\nu) + j\tilde{U}_{mn}(\nu) \rangle_\nu + \langle D^R_m (\nu) e^{2j\phi_m}r^{LL}_{mn, {\rm cal}} (\nu) \rangle_\nu \nonumber \\ &+& \langle D^{L*}_n(\nu)e^{2j\phi_n}r^{RR}_{mn, {\rm cal}} (\nu) \rangle_\nu \nonumber \\ &+& \langle D^R_{m}(\nu)D^{L*}_{n}(\nu)e^{2j(\phi_m+\phi_n)}(\tilde{Q}_{mn}(\nu) - j\tilde{U}_{mn}(\nu)) \rangle_\nu, \nonumber \\
\end{eqnarray}
where $\langle \cdots \rangle_\nu$ indicates averaging over frequency. The source's emission is not expected to change significantly within the frequency range. As a result, $\tilde{Q}_{mn}$, $\tilde{U}_{mn}$, $r^{LL}_{mn, {\rm cal}}$, and $r^{RR}_{mn, {\rm cal}}$ can be approximated as a constant if we ignore the small variations in these quantities with frequency caused by the different $(u,v)$ coverage. Thus, Equation~\ref{eq:avgmodel1} can be rewritten as follows:
\begin{eqnarray}
\label{eq:avgmodel2}
\langle r^{RL}_{mn}(\nu) \rangle_\nu &=& {\rm const} + \langle D^R_m (\nu) \rangle_\nu \times {\rm const} + \langle D^{L*}_n(\nu) \rangle_\nu \times {\rm const} \nonumber \\ &+& \langle D^R_{m}(\nu)D^{L*}_{n}(\nu) \rangle_\nu \times {\rm const}.
\end{eqnarray}
In the absence of the second-order term, the D-terms derived using frequency-averaged visibilities must equal the average of the spectra of the true D-terms. In other words, if the second-order term is not present, correcting for D-terms by using frequency-averaged data is equivalent to correcting for D-terms for each frequency channel and then performing frequency averaging. Due to the second-order terms, however, the two cases differ as $\langle D^R_{m}(\nu)D^{L*}_{n}(\nu) \rangle_\nu \neq \langle D^R_{m}(\nu) \rangle_\nu \langle D^{L*}_{n}(\nu) \rangle_\nu$. There is an exception to this relationship when the D-terms are constant over frequency, which we do not take into account in this paper. Thus, if the D-term amplitudes are small, the effect of calibration of frequency-dependent instrumental polarization will be small.

\section{Test using synthetic data with antenna gain corruptions}
\label{appendix:synthetic}

A synthetic data set with antenna gain matrices assumed to be identity matrices was used in Section~\ref{sec:synthetic} to verify GPCAL's frequency-dependent leakage calibration method. Any realistic VLBI data, however, will contain antenna gain corruptions. In this Appendix, we test the method with a synthetic data set containing gain corruptions.

We generated a synthetic data set in accordance with Section~\ref{sec:synthetic}, but we included the corruption of antenna gain in the data set. Antenna gains were assumed to be constant over frequency and to vary from scan to scan, but to remain constant within a scan. The gain amplitudes for each station are randomly drawn from a Gaussian distribution with a standard deviation of 0.05. An overall shift in the gain amplitudes from unity has been introduced by a uniform random value between $-0.1$ and $0.1$ for each station. Following the synthetic data generation conducted for the EHT analysis \citep[e.g.,][]{EHT2019d}, a random station phase is adopted for each scan to simulate the absence of absolute phase.

\begin{figure*}[t!]
\centering
\includegraphics[width = \textwidth]{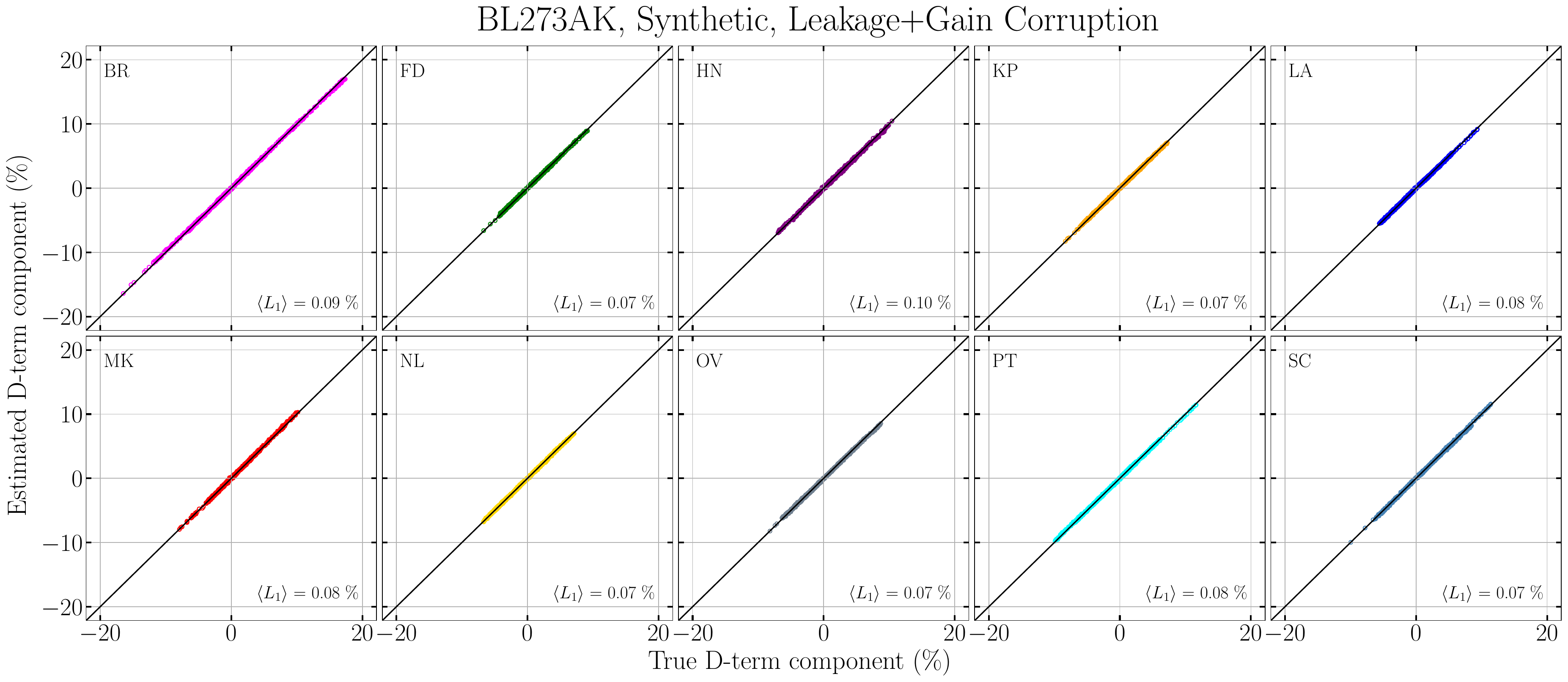}
\caption{Same as Figure~\ref{fig:dsyn1to1} but for the synthetic data including both frequency-dependent polarimetric leakage and antenna gain corruptions. \label{fig:gdsyn1to1}}
\end{figure*}

\begin{figure*}[t!]
\centering
\includegraphics[width = \textwidth]{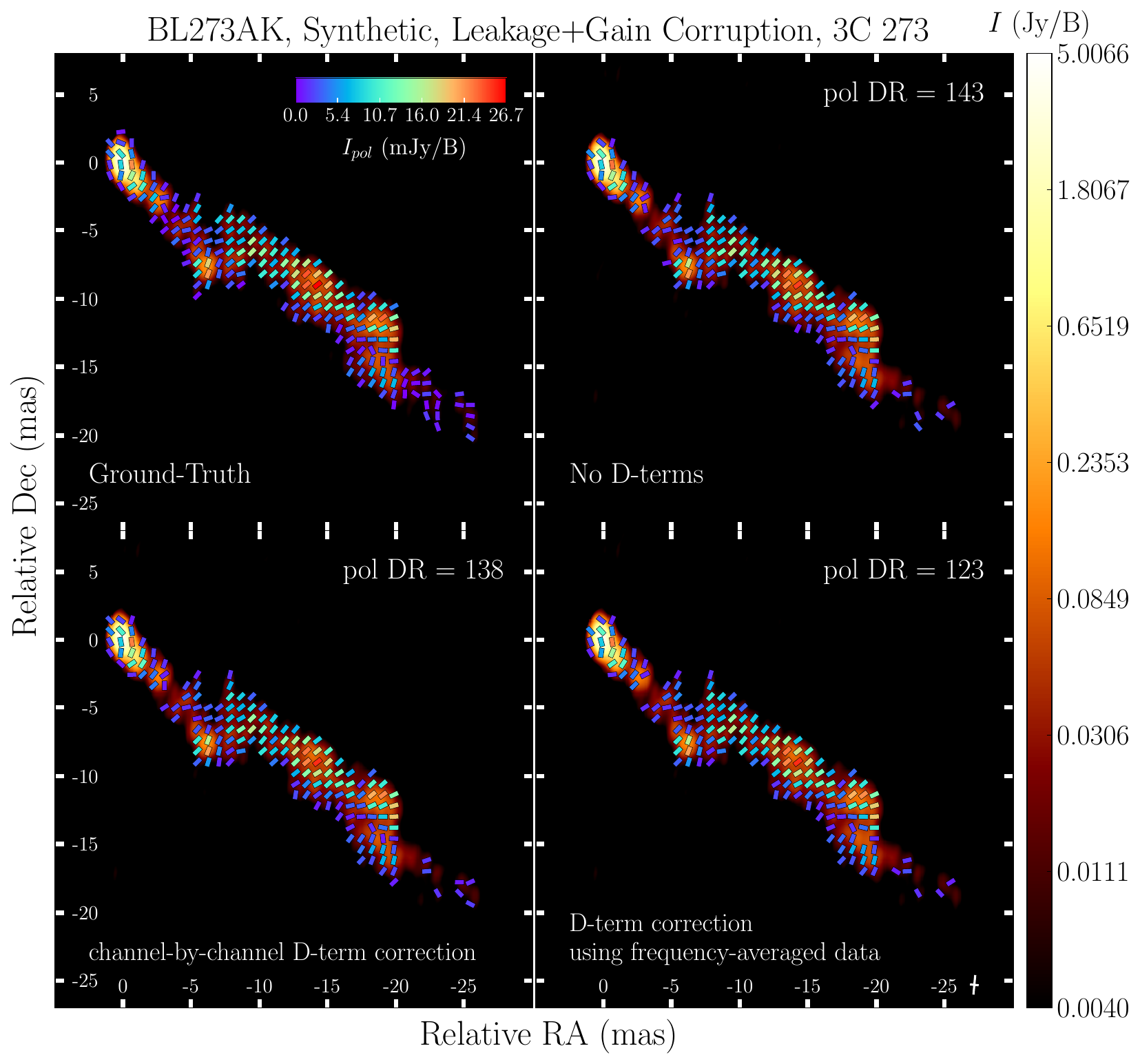}
\caption{Same as Figure~\ref{fig:dsynimage} but for the synthetic data including both frequency-dependent polarimetric leakage and antenna gain corruptions. \label{fig:gdsynimage}}
\end{figure*}

We derived frequency-dependent D-terms based on the synthetic data sets, as described in Section~\ref{sec:synthetic}. Due to the presence of gain corruptions in this data set, we performed iterative CLEAN and self-calibration on Stokes I data prior to frequency-dependent leakage calibration. The reconstructed D-term components (real and imaginary parts) are compared to the ground truth components in Figure~\ref{fig:gdsyn1to1}. For all VLBA stations, we were able to reproduce the ground-truth D-terms with an accuracy of the average $L_1$ norm that was slightly higher than the accuracy obtained in the synthetic data test without gain corruption (Section~\ref{sec:synthetic}). In Figure~\ref{fig:gdsynimage}, we present the linear polarization images of 3C 273 obtained from the data without including any leakage (``No D-terms"; top right), as well as the images obtained from the leakage corrupted data with (bottom left) and without (bottom right) frequency-dependent leakage correction. 

\begin{figure*}[t!]
\centering
\includegraphics[width = \textwidth]{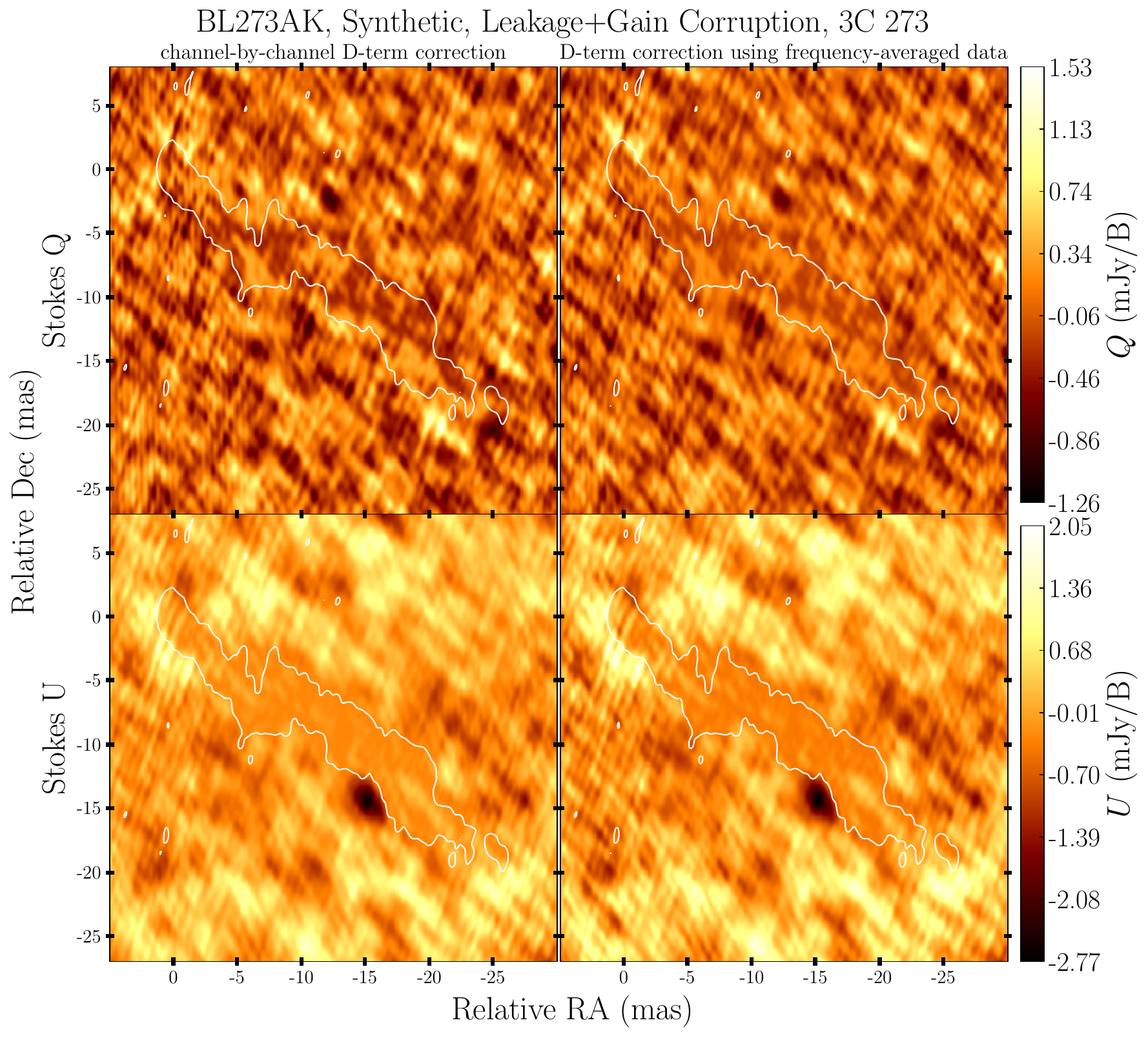}
\caption{Same as Figure~\ref{fig:dsyndirty} but for the synthetic data including both frequency-dependent polarimetric leakage and antenna gain corruptions. \label{fig:gdsyndirty}}
\end{figure*}

The DR of the image obtained from the data without including D-terms is already much lower than the DR obtained from the data without inclusion of gain corruptions (the top right panel in Figure~\ref{fig:dsynimage}). This value is similar to that of the DR values for images with and without frequency-dependent leakage corrections from gain corrupted synthetic data. We present the dirty images of the residual visibilities for Stokes $Q$ and $U$ in Figure~\ref{fig:gdsyndirty}. The noise levels did not exhibit significant improvement with the frequency-dependent leakage correction for this synthetic data, which contrasts with the results observed for the synthetic data without gain corruptions (Section~\ref{sec:synthetic}). It appears that the gains could not be perfectly calibrated, which has a significant impact on the quality of the polarization images. Relative to the degradation caused by residual gain errors, the quality improvement due to frequency-dependent leakage correction is marginal. 

The present result is not surprising since the major contribution of D-terms to cross-hand visibilities is linear, as outlined in Appendix~\ref{appendix:linear}. The distortion in cross-hand visibilities caused by averaging data over frequency prior to correcting the frequency-dependent D-terms introduced by the second-order terms in Equation~\ref{eq:fitmodel} is expected to be small if the amplitudes of the D-terms are small. It is possible that this distortion may be dominated by other systematic errors, such as residual gain errors. It is expected that realistic VLBI data will contain gain errors and this is probably why we did not see much improvement in the image DR for the real VLBA data sets in Section~\ref{sec:application}.

\bibliography{AAS44256R2}{}
\bibliographystyle{aasjournal}

\end{document}